\documentclass[12pt]{article}
\usepackage{graphics,epsfig}

\def\appendix#1{
  \addtocounter{section}{1}
 \setcounter{equation}{0}
  \renewcommand{\thesection}{\Alph{section}}
 \section*{Appendix \thesection\protect\indent \parbox[t]{11.715cm} {#1}}
  \addcontentsline{toc}{section}{Appendix \thesection\ \ \ #1}
  }

\def\bea{\begin{eqnarray}}
\def\eea{\end{eqnarray}}
\def\be{\begin{equation}}
\def\ee{\end{equation}}

\newcommand{\Tr}{\mathop{\mathrm{Tr}}\nolimits}
\newcommand{\lvert}{\mathopen{|}}
\newcommand{\rvert}{\mathclose{|}}

\def\d{\partial}
\def\D{\delta}
\def\dd{^{\dagger}}
\def\gym{g}

\newcommand{\rf}[1]{(\ref{#1})}
\newcommand{\non}{\nonumber \\*}
\def\N{{\cal N}=4}
\def\ads{$AdS_5\times S_5$}
\def\al{\alpha}
\def\l{\lambda}
\def\w{{\cal W}}
\def\ww{{\cal W}_{g\rightarrow 0}}
\def\ws{{\cal W}_{g\rightarrow \infty}}

\begin{document}
\begin{titlepage}
\begin{flushright}
ITEP--TH--25/99
\end{flushright}
\vspace{1.5cm}

\begin{center}
{\LARGE Structure of the Electric Flux} 
\\[.5cm]
{\LARGE in $\N$ Supersymmetric Yang-Mills Theory}\\
\vspace{1.9cm}
{\large J.~Erickson\footnote{
Department
of Physics and Astronomy, University of British Columbia, 6224 Agricultural
Road, Vancouver, British Columbia, Canada V6T 1Z1.}, G.W.~Semenoff$\,^{*}$
and K.~Zarembo$^{*,}$\footnote{
Institute of Theoretical and Experimental Physics,
B. Cheremushkinskaya 25, 117259 Moscow, Russia.\hfill\vskip 5pt
\hspace*{0.37em}E-mail: \texttt{erickson@nbi.dk}, \texttt{semenoff@nbi.dk}, \texttt{zarembo@itep.ru}}}
\\
\vspace{24pt}
{\it The Niels Bohr Institute\\
Blegdamsvej 17\\
DK-2100 Copenhagen \O\\
Denmark}
\end{center}
\vskip 2 cm
\begin{abstract}
Correlators of Wilson loop operators with ${\cal O}_4= {\rm
Tr}(F_{\mu\nu}^2+\cdots)$ are computed in ${\cal N}=4$
super-Yang-Mills theory using the AdS/CFT correspondence.  The results
are compared with the leading order perturbative computations. As a
consequence of conformal invariance, these correlators have identical
forms in the weak and strong coupling limits for circular loops. They
are essentially different for contours not protected by conformal
symmetry.
\end{abstract}

\end{titlepage}

\setcounter{page}{2}


The conjectured duality of $\N$, $D=4$ super-Yang-Mills theory (SYM)
to supergravity on \ads~gives rise to the possibility of obtaining
quantitative information about the former theory in the limit of
infinite 't~Hooft coupling \cite{Mal97,Gub98,Wit98,Aha99}. Though this
is a regime which is inaccessible to conventional perturbative field
theory methods, some of the results obtained from the AdS/CFT
correspondence can be tested by non-renormalization theorems which
protect certain correlation functions from gaining radiative
corrections
\cite{Aha99}\footnote{Various non-renormalization theorems for
$\N$ SYM are discussed in
\cite{Gub97,Ans97,Fre98,Lee98,DHo98,Int98,Gon99,Int99,Ede99,Pet99,Liu99}.}.
The quantities not protected by supersymmetry probe the genuine
strong-coupling dynamics of the SYM theory.  An example of such
a quantity is the interaction potential between static charges.  The
Coulomb form of the potential is dictated by the conformal invariance
of $\N$ SYM, but the strength of the interaction is not fixed by any
symmetry and appears to be proportional to $\sqrt{g^2N}$ at strong
coupling
\cite{Mal98,Rey98} in contrast to the weak-coupling $O(g^2N)$ behavior.

The purpose of the present paper is to study the structure of the
electric field induced by static charges in the vacuum. The static charges
are introduced by a Wilson loop. The one appropriate 
for supergravity calculations corresponds to
the phase factor acquired by an infinitely heavy W-boson propagating in
background gauge and scalar fields \cite{Mal98,Dru99}:
\be
\label{wl}
W(C)=\frac{1}{N}\,\Tr {{\rm P}}\exp\oint_C d\tau\,(iA_\mu\dot{x}_\mu
+\Phi_i\dot{y}_i), 
\ee 
where the $A_\mu$ are $U(N)$ gauge fields and the
$\Phi_i$ are scalars in the adjoint representation, both represented by
Hermitian matrices. Moreover,
$\dot{y}_i=\sqrt{\dot{x}^2}\theta_i$ where $\theta_i$ is a point on the
five-dimensional sphere: $\theta^2=1$.  

The form of the coupling to
scalars in \rf{wl} is natural from several points of view
\cite{Dru99,Mak88}.  In particular, this Wilson loop operator is
annihilated by half of the supercharges if $C$ is a straight line
\cite{Dru99}.  On the other hand, closed Wilson loops are generally
not supersymmetric.  A circular loop, however, is a special case since
it can be mapped to a straight line by a conformal transformation. The
straight line Wilson loop operator commutes with half of the
supersymmetry generators.  Thus, circular loops are annihilated by a
combination of supercharges twisted by conformal
transformations\footnote{In discussing supersymmetry, it is
convenient to consider $\N$ SYM theory as a reduction of
ten-dimensional ${\cal N}=1$ theory. Then the supercharges are the
components of a ten-dimensional Majorana-Weyl spinor $Q_\al$. The
gauge potentials and the scalar fields form a ten-dimensional vector:
$A_M=(A_\mu,\Phi_i)$. The supercharges act on gauge potentials as
$[Q_\al,A_M]=\frac{i}{2}(\Gamma_M\Psi)_\al$, where $\Gamma_M$ are
Dirac matrices and $\Psi$ is the ten-dimensional Majorana-Weyl
fermion, the reduction of which to four dimensions gives the fermionic
content of $\N$ SYM theory. The supersymmetry transformation of a
Wilson loop produces degenerate combination of the Dirac matrices
\cite{Dru99}: $\bar{\epsilon}\dot{X}^M\Gamma_M \Psi$,
$X^M=(x^\mu,iy^i)$, since $\dot{X}^2=0$. An appropriate choice of
$\epsilon$ turns the variation of a Wilson loop to zero if $\dot{X}^M$
is constant, that is, if $C$ is a straight line.}:
\be
\tilde{Q}_\al=\Omega(C)\dd Q_\al \Omega(C),
\ee
where $\Omega(C)$ is an unitary operator implementing a conformal
transformation that maps $C$ to a line. Since conformal invariance is
an exact symmetry of the $\N$ SYM vacuum, $\Omega |0\rangle=|0\rangle$
and the twisted supercharges annihilate the vacuum:
$\tilde{Q}_\al|0\rangle=0$, along with the original supersymmetry
generators $Q_\al$.  If the Wilson loop has a different geometry,
generically there is no residual supersymmetry.  This is the case for a loop consisting of two infinite anti-parallel lines; this loop corresponds to a static electric dipole.


The structure of electric field induced by the charges can be probed by
local gauge invariant operators, like $\Tr F^2_{\mu\nu}$. On the supergravity 
side, 
\be
{\cal O}_4(x)=
\frac{1}{2g^2}\Tr\left\{F_{\mu\nu}^2+[\Phi_i,\Phi_j]^2+
\bar\psi\Gamma^i[\Phi_i,\psi]\right\}
\ee
couples to the dilaton. We consider the 
correlator of the Wilson loop with ${\cal O}_4(x)$:
\be\label{defw}
\w(C,x)=-\,\frac{\bigl\langle W(C)\,\,\frac{1}{2g^2}
\Tr \bigl(F_{\mu\nu}^2(x)+\cdots\bigr)\bigr\rangle_{\rm conn}}
{\left\langle W(C)\right\rangle},
\label{corr}
\ee
where the dots denote the scalar and the fermionic terms required to
put the operator in the short multiplet of $\N$ SUSY.
The correlator \rf{defw}
probes the local electric field induced 
by non-dy\-na\-mi\-cal charges whose trajectories form the loop $C$.
This correlator has been discussed in the context of the AdS/CFT 
correspondence in the limiting case when the distance from the point $x$ to
the loop $C$ is large and the Wilson loop can be expanded in local 
operators \cite{Ber98}. The correlator \rf{defw} was also used to
demonstrate that, in the finite-temperature, confining phase of
strongly coupled $\N$ SYM,  the classical string in the bulk
is projected onto an electric flux tube, stretched between static
charges, which has thickness on the order of
the correlation length \cite{Dan98}\footnote{More applications are
discussed in \cite{Cal99}.}. In the conformal
theory, the correlator must fall-off as a power of the distance from
the loop. At weak coupling, the correlator decreases with distance as
$1/d^6$, since the the classical electric field of the dipole falls off
as $1/d^3$. We shall show that the correlator obeys a different power law
at strong coupling.  

At strong coupling, correlators in the SYM theory are described by classical
supergravity on \ads. We choose units in which the radius of $AdS_5$ is
unity. The metric on \ads~in these units is
\be
ds^2=\frac{1}{z^2}\left(dz^2+dx_\mu dx_\mu\right)+d\Omega_5^2.
\ee
The Wilson loop on the boundary of $AdS_5$ at $z=0$ creates the classical 
string in the bulk; it is the surface of minimal area 
bounded by the loop \cite{Mal98,Rey98}.
The correlation function
\rf{defw} is given by the dilaton propagator with one end 
attached to the point $x$ and the other integrated over the minimal
surface spanning the loop $C$ \cite{Ber98}:
\be
\ws\propto\sqrt{\gym^2 N}\int d^2\sigma\,
\sqrt{\det G}\,\frac{Z^4}{\bigl[Z^2+\lvert \vec{X}-\vec{x}\rvert^2\bigr]^4},
\label{wloopG}
\ee
where $G_{ab}=g_{MN}\d_a Y^M\d_b Y^N$
is the metric induced on the minimal surface. The surface is
parameterized by coordinates $Y^M=\Bigl(Z(\sigma),X^\mu(\sigma)\Bigr)$
in $AdS_5$. 

At weak coupling, the correlator \rf{defw} is given by the
lowest-order Feynman diagram:
\bea
\ww&=&\frac{g^2N}{4}\,\oint_C dy_\mu\,\oint_C dz_\nu\left[
\D_{\mu\nu}\,\,\frac{\d}{\d x_\l}\,\frac{1}{4\pi^2\lvert\vec{x}-\vec{y}\rvert^2}
\,\,\frac{\d}{\d x_\l}\,\frac{1}{4\pi^2\lvert\vec{x}-\vec{z}\rvert^2}
\right.\non && \qquad\left.
-\frac{\d}{\d x_\nu}\,\frac{1}{4\pi^2\lvert\vec{x}-\vec{y}\rvert^2}
\,\,\frac{\d}{\d x_\mu}\,\frac{1}{4\pi^2\lvert\vec{x}-\vec{z}\rvert^2}
\right].
\eea

We start with the supersymmetry preserving circular loop. 
At weak coupling, we find:
\be\label{weak1}
\ww=\frac{g^2N}{4\pi^2}\frac{R^4}{\bigl[(y^2+r^2-R^2)^2+4R^2y^2\bigr]^2}\,,
\ee
where $R$ is the radius of the loop, $r$ is the radial coordinate of the
point $x$, and $y$ is the distance from $x$ to 
the plane of the loop.
This expression coincides with the result of the strong-coupling
supergravity calculation given in \cite{Ber98}: 
\be
\ws\propto\sqrt{\gym^2 N}\frac{R^4}{\bigl[(y^2+r^2-R^2)^2+4R^2y^2\bigr]^2}.
\ee
The functional form of the correlator of $\Tr (F^2+\cdots)$ with
the circular loop appears not to be renormalized. Actually, this 
non-renormalization property 
is a consequence of conformal symmetry. Although the scale
invariance only fixes the above correlator up to a function of two 
dimensionless ratios of the parameters $r$, $R$ and $y$, the
conformal transformation which maps the circle to the line reduces
the number parameters to one, the distance $d$ between the point and
the line. The correlator of $\Tr (F^2_{\mu\nu}+\cdots)$ with
the Wilson line is determined by the dimension of the dilaton 
operator and is proportional $1/d^4$. Transforming the line
to the circle we get \rf{weak1} with an overall coefficient which can 
depend only on the coupling. 
The distinguished role of the supersymmetry preserving
circular loops is reflected
in the fact that minimal surface in $AdS_5$ bounded by it
has a very simple form 
\cite{Dru99,Ber98}. The minimal surfaces bounded by contours which do not have such nice conformal properties are generically more complicated
\cite{Mal98,Rey98,Zar99,Dru99}. We will find that for such contours the electric field profile probed by the correlator \rf{defw} is renormalized in a non-trivial way.

We consider the loop composed of antiparallel infinite lines.
The leading-order perturbative result for the correlator with the
dilaton operator
in this case is
\bea\label{wpert}
\ww&=&\frac{g^2N}{64\pi^2}\Biggl\{
\frac{1}{\bigl[y^2+(d-L/2)^2\bigr]^2} -
2\frac{y^2+d^2-L^2/4}{\bigl[y^2+(d-L/2)^2\bigr]^{3/2}
\bigl[y^2+(d+L/2)^2\bigr]^{3/2}} \non
&&\qquad\qquad+\frac{1}{\bigl[y^2+(d+L/2)^2\bigr]^2}\Biggr\}
\non
&=&\frac{g^2N}{4}
\left[\frac{\d}{\d x_i}\left(\frac{1}{4\pi|{\bf x}-{\bf L}/2|}
-\frac{1}{4\pi|{\bf x}+{\bf L}/2|}\right)\right]^2,
\eea 
where $L$ is the distance between the lines; $d$ is the distance, in
the plane of the lines, between the point $x$ and the line 
midway between the two lines; and $y$ is the distance between 
$x$ and the plane of the lines: ${\bf L}=(0,L,0,0)$, ${\bf x}=(0,d,y,0)$.

An expression for the minimal surface whose boundary is the two lines
is given in \cite{Mal98}:
\be
x=Z_0\int_1^{Z_0/Z}\frac{du}{u^2\sqrt{u^4-1}},
\ee
with
\be
Z_0=\frac{L}{2}\biggl[\int_1^\infty\frac{du}{u^2\sqrt{u^4-1}}\biggr]^{-1}
=\frac{L}{2}\frac{\Gamma(1/4)^2}{\sqrt{2}\pi^{3/2}}.
\ee
From \rf{wloopG} we find:
\bea
\ws&\propto&\sqrt{g^2N}\,Z_0^2\int_{-\infty}^\infty 
dt\int_{-L/2}^{L/2} dx 
\frac{1}{[Z^2+y^2+(x-d)^2+t^2]^4}
\non
&\propto&
\sqrt{g^2N}L^2\int_{-L/2}^{L/2} dx\,\frac{1}{[Z^2+y^2+(x-d)^2]^{7/2}},
\label{apmalda}
\eea 
where $y$, $L$, and $d$ are as before. 

\begin{figure}
\begin{center}
\epsfig{file=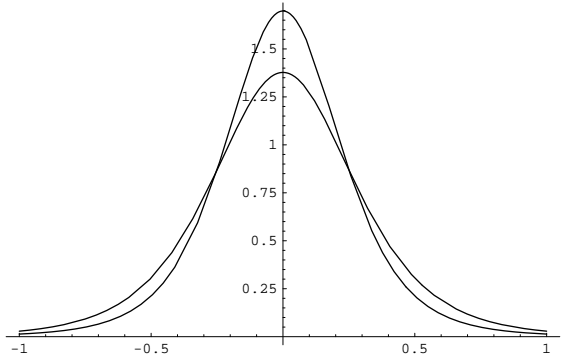,width=3in}
\end{center}
\caption{$\ws$ (the shorter peak) and $\ww$ (the taller peak), drawn
as a function of $y$ with $L=1$ and $d=0$. The heights are normalized
so that the area under each curve is equal to unity.  Note that
asymptotically the $\ws$ falloff is actually faster ($y^{-7}$ versus
$y^{-6}$).}
\label{fig:wilson}
\end{figure}

This result is clearly different than that of the perturbative
calculation. In particular, we can consider the asymptotic behavior of
each result when the distance to the point $x$ is much larger than the
separation between the charges $L$.  Expanding (\ref{apmalda}) in $L$
gives
\be
\ws\propto\frac{\sqrt{g^2N}\,L^3}{(y^2+d^2)^{7/2}}+O(L^5).
\ee
The perturbative correlator \rf{wpert} is just the field strength
of electric dipole squared and is proportional to $L^2$ for small $L$:
\be
\ww=\frac{g^2N}{64\pi^2}\,\frac{(4d^2+y^2)L^2}{(y^2+d^2)^4}+O(L^4).
\ee
Thus, the two correlators obey different power laws at large $d$, with
$\ws$ falling off as $d^{-7}$ and $\ww$ falling off as
$d^{-6}$. Figure~\ref{fig:wilson} shows the form of the correlator \rf{defw}, as a function of $y$,
at strong and weak coupling, for $d=0$.


However, the leading singularity as the point $x$ approaches the loop
obeys the same power law in the weak and the strong coupling
limits. In both cases, the correlator behaves as $\epsilon^{-4}$,
where $\epsilon=d-L/2$ is the distance from the loop, as can be
checked from \rf{wpert} and \rf{apmalda}. This is to be expected as the dimension of the dilaton operator is not renormalized.

To conclude, we have calculated correlators of Wilson loops with the
dimension four chiral
operator $\Tr (F^2+\cdots)$ in $\N$ SYM. We have compared the perturbative
expression with the strong coupling result predicted by the AdS/CFT correspondence.  The correlator for a circular loop has
the same functional form in both regimes, whereas the answer is
completely different for antiparallel Wilson lines. Even the
asymptotic form of the correlator at infinity  appears to be modified.

Finally, we observe that the coupling constant dependence of $\w(C,x)$
at large $N$ is natural and is related to the strong coupling limit of
the Wilson loop expectation value.  Indeed, ${\cal O}_4$ is related to
the action density of super-Yang-Mills theory if we add terms which
are a total divergence, and use the equations of motion for
the scalar and fermion fields.  Thus, the integral of (\ref{corr})
over the coordinate point $x$ is related to the derivative of the
expectation value of the Wilson loop by the coupling constant,
\be
\int d^4x\,\w(C,x)=g^2\frac{\partial}{\partial g^2}\ln W(C)
\ee
Also, we see from (\ref{wloopG}) that, in the $g\rightarrow\infty$
limit, this integral of $\w(C,x)$ is equal to $\pi^2/6$ times the
minimal area of the surface.  Thus, if the Wilson loop correlator at
strong coupling is $\ln W(C)=\sqrt{4\pi g_{YM}^2N}$ then we would
expect that $\w(C,x)=3\sqrt{4\pi g_{YM}^2N}/\pi^2$.

\section*{Acknowledgments}

We are grateful to Y.~Makeenko  and M.~van~Raamsdonk for discussions and
to J.~Ambj{\o}rn for hospitality at the Niels Bohr Institute.
The work of KZ was supported by NATO Science Fellowship and, in part, by
 INTAS grant 96-0524,
 RFFI grant 97-02-17927
 and grant 96-15-96455 for the promotion of scientific schools.  GWS was
supported by NSERC of Canada, the Niels Bohr Fund of Denmark and NATO
CRG 970561.


\end{document}